\documentstyle[12pt,texdraw]{article}
\input texdraw
\def\ftoday{{\sl  \number\day \space\ifcase\month
\or Janvier\or F\'evrier\or Mars\or avril\or Mai
\or Juin\or Juillet\or Ao\^ut\or Septembre\or Octobre
\or Novembre \or D\'ecembre\fi
\space  \number\year}}
\newcommand{\journal}[4]{{\em #1~}#2\,(19#3)\,#4;}

\newcommand{\hpa}{\journal {Helv. Phys. Acta}}

\newcommand{\ijmp}{\journal {Int. J. Mod. Phys.}}

\newcommand{\np}{\journal {Nucl. Phys.}}
\newcommand{\pl}{\journal {Phys. Lett.}}
\newcommand{\mpl}{\journal {Mod. Phys. Lett.}}
\newcommand{\prep}{\journal {Phys. Rep.}}

\makeatletter
\@addtoreset{equation}{section}
\makeatother

\newcommand{\es}{\\[3mm]}

\renewcommand{\d}{\delta}         \newcommand{\D}{\Delta}
\newcommand{\e}{\varepsilon}

\newcommand{\la}{\lambda}        
\newcommand{\m}{\mu}
\newcommand{\n}{\nu}
\newcommand{\om}{\omega}         \newcommand{\OM}{\Omega}
\newcommand{\p}{\psi}             
\def \r {\rho}
\newcommand{\s}{\sigma}           \renewcommand{\S}{\Sigma}
\newcommand{\th}{\theta}         
           
\newcommand{\vf}{{\varphi}}
\newcommand{\x}{\xi}              \newcommand{\X}{\Xi}

\newcommand{\BB}{{\cal B}}

\newcommand{\Sla}{\raise.15ex\hbox{$/$}\kern -.70em D}

\newcommand{\lp}{\left(}\newcommand{\rp}{\right)}
\newcommand{\lc}{\left[}\newcommand{\rc}{\right]}
\newcommand{\lac}{\left\{}\newcommand{\rac}{\right\}}

\newcommand{\complex}{{\kern .1em {\raise .47ex
\hbox {$\scriptscriptstyle |$}}
    \kern -.4em {\rm C}}}
\newcommand{\real}{{{\rm I} \kern -.19em {\rm R}}}
\newcommand{\rational}{{\kern .1em {\raise .47ex
\hbox{$\scripscriptstyle |$}}
    \kern -.35em {\rm Q}}}
\renewcommand{\natural}{{\vrule height 1.6ex width
.05em depth 0ex \kern -.35em {\rm N}}}

\newcommand{\tr}{{\rm {Tr} \,}}

\newcommand{\pa}{\partial}

\newcommand{\fud}[2]{{\frac{\delta #1}{\delta #2}}}

\newcommand{\dfrac}[2]{{\displaystyle{\frac{#1}{#2}}}}
\newcommand{\dsum}[2]{\displaystyle{\sum_{#1}^{#2}}}
\newcommand{\dint}{\displaystyle{\int}}

\newcommand{\ie}{{{\em i.e.},\ }}

\newcommand{\sla}{\raise.15ex\hbox{$/$}\kern -.57em}
\newcommand{\twiddle}{\lower.9ex\rlap{$\kern -.1em\scriptstyle\sim$}}

\newcommand{\vev}[1]{\left\langle {#1}\right\rangle}
\newcommand{\equ}[1]{(\ref{#1})}
\newcommand{\eq}{\begin{equation}}
\newcommand{\eqn}[1]{\label{#1}\end{equation}}
\newcommand{\eea}{\end{eqnarray}}
\newcommand{\eqa}{\begin{eqnarray}}
\newcommand{\eqan}[1]{\label{#1}\end{eqnarray}}
\newcommand{\ba}{\begin{array}}
\newcommand{\ea}{\end{array}}
\newcommand{\eqac}{\begin{equation}\begin{array}{rcl}}
\newcommand{\eqacn}[1]{\end{array}\label{#1}\end{equation}}

\renewcommand{\=}{&=&} 
\renewcommand{\title}[1]{\null\vspace{25mm}

\noindent{\Large{\bf #1}}\vspace{10mm}

}
\newcommand{\authors}[1]{\noindent{\large #1}\vspace{20mm}

}
\newcommand{\address}[1]{{\center{\noindent #1\vspace{10mm}}

}}
\renewcommand{\abstract}[1]{\vspace{17mm}

\noindent{\small{\em Abstract.} #1}\vspace{2mm}

}
\newcommand{\dprod}[2]{\displaystyle{\prod_{#1}^{#2}}}
\newcommand{\zc}{Z_{\rm c}}
\newcommand{\xtr}{x^{\rm tr}}
\newcommand{\ytr}{y^{\rm tr}}

\newcommand{\nes}{\nonumber\es}
\def\+{\!\!\!&&\!\!\!+~}

\newcommand{\mayo}{
\fcir f:0 r:0.04
\rlvec(1 0)
  \bsegment
     \linewd 0.04
     \lvec(0 0.5)  \rlvec(1 0)  \rlvec(0 -1)
     \rlvec(-1 0)   \rlvec(0 0.5)
     \textref h:C v:C \htext(0.5 0){$Z_+$}
   \esegment
\rmove(1 0.35)  \rlvec(0.7 0)  \fcir f:0 r:0.04
\rmove(-0.7 -0.7)   \rlvec(0.7 0)  \fcir f:0 r:0.04
\rmove(0 0.25)    \fcir f:0 r:0.02
\rmove(0 0.1)     \fcir f:0 r:0.02
\rmove(0 0.1)     \fcir f:0 r:0.02 }
\begin{document}
\begin{titlepage}
\begin{center}
\hspace*{\fill}{{\normalsize \begin{tabular}{l}
                              {\sf TUW 95-14}\\
                              {\sf  UGVA---DPT 1995/06--895}\\
                              {\sf hep-th/9506208}
                         \end{tabular}   }}

\title{Five-Dimensional BF Theory and \\[2mm]
               Four-Dimensional Current Algebra}

\authors{S. Emery$^{1\ \vdash}$, H. Jirari$^\dashv$,
               O. Piguet$^{2\ \dashv}$}

\address{$^\vdash$ Institut f\"ur Theoretische Physik,
                          Technische Universit\"at Wien\\
      Wiedner Hauptstra\ss e 8-10, A-1040 Wien (Austria)\\[3mm]
        $^\dashv$ D\'epartement de Physique Th\'eorique,
                          Universit\'e de Gen\`eve\\
                   CH -- 1211 Gen\`eve 4 (Switzerland)}

\end{center}
\footnotetext[1]{On leave from the D\'epartement de Physique Th\'eorique,
  Universit\'e de Gen\`eve. Supported by the ``Fonds Turrettini'' and
  the ``Fonds F. Wurth''}
\footnotetext[2]{Supported in part
                 by the Swiss National Science Foundation.}
\abstract{
We consider the relation between the five-dimensional
 BF model and a four-dimensional
local current algebra from the point of view of perturbative
local quantum
field theory. We use an axial gauge fixing procedure and show that
it allows for a well defined theory which actually can be solved exactly.
}

\end{titlepage}

\section{Introduction}

The relation between Topological Field Theories~\cite{bbrt} and local
Field Theories is well established for the three dimensional
Chern-Simons (CS) model. Indeed,
it is well known that the restriction of the latter on a
two dimensional plane leads to physical observables, namely the
two dimensional conserved chiral currents generating the Kac-Moody
algebra of the Wess-Zumino type~\cite{csbord}. In a previous
paper~\cite{emery-piguet}, we exhibit this fact using a very general
procedure.   More precisely, after having defined the model on a
manifold with boundary, we chose to implement the effects of the
latter by means of two requirements:
a decoupling condition which forbids the existence of any
interactions through the boundary, and a locality condition which
states that away from it, the theory is the same as the one without
boundary. It is remarkable that such conditions can be implemented
directly at the level of the generating functional for the connected
Green function.   Therefore, it avoids the problem of dealing with
surface terms which are {\it a priori} ill defined products of
distributions at the same point and which
would need to be regularized.

In~\cite{emery-piguet} an axial gauge was chosen,
which is a natural choice when
considering a plane boundary in a plane space-time.
Such a gauge fixing is however incomplete, invariance under residual
gauge transformations being left.
It is precisely the Ward identity expressing this residual invariance
which has been interpreted as a current algebra on the boundary.

In this paper, we will apply the procedure
of~\cite{emery-piguet} in higher
dimensions. More precisely, we shall start with a
five-dimensional BF model
defined on a  manifold with a boundary of dimension 4, and look
at the consequences of the Ward identities which correspond to the
residual gauge invariances. At this point, let us note that
the symmetry content of the BF system is greater than that of the
CS theory. Indeed, such models are known to exhibit reducible
symmetries~\cite{bbrt,luc-pig-sor}. Therefore the question is to know
whether such a difference breaks the full procedure. The answer
is negative. The present theory
possesses a four-dimensional current algebra which lies
on the boundary. This algebra is the one which is generated
through the residual Ward identity of the Yang-Mills
symmetry. The
one which corresponds to the reducible symmetry contains a hard
breaking and cannot be interpreted as a current algebra.

It should be noticed that the derivation of the  Ward identities
for the residual gauge invariances suffers from
an infrared (IR) problem, linked to the bad long distance
behaviour associated with the particular geometry implemented, and
particularly with the gauge condition given by
the vanishing of the gauge field
components which are orthogonal to the boundary plane.
Therefore, one will have to make use of an infrared
regularized gauge condition.

The paper is organized as follows. We first review some general
facts about the
five-dimensional BF theory in the axial gauge and the procedure
followed for describing the effects of the boundary. Then we
compute the propagators and the $N$-point Green
functions as the general solution of the field equations.
Finally we prove the
residual Ward identity leading to the algebra of
currents living on the boundary considered as a four-dimensional
space-time. It is only for this last point that the infrared
regularization is introduced, since the Green functions exist without
such a regularization.
We finish with some conclusions.


\section{Five-Dimensional BF Theory in the Axial \break Gauge}
The classical action of the BF model in 5 dimensions
reads\footnote{{\bf Conventions}:
$\m,\n,\cdots = 0,1,2,3,4\ $,
      $g_{\m\n}= {\rm diag}(1,-1,-1,-1,-1)\ $,
$\e^{\m\n\r\s\tau}=\e_{\m\n\r\s\tau}=\e^{[\m\n\r\s\tau]}$,
$\e_{01234}=1$.}
\eq
\Sigma_{\rm BF}=\dint \tr\lp B\wedge F\rp
=\frac{1}{2\cdot3!}\dint d^5x\ \e^{\m\n\r\s\tau}
             \tr\lp B_{\m\n\r}F_{\s\tau}\rp
\eqn{action}
where $B$ is a three form and  $F= dA+\frac{1}{2}[A,A]$ is the
field strength of the gauge connection $A$.
These fields, as well as all the one's encountered throughout this
paper belong to the adjoint
representation and are written as Lie algebra matrices
$\vf (x) = \vf (x)^a \tau_a$ with
\[
[\tau_a,\tau_b]= f_{ab}{}^c \tau_c,\qquad \tr(\tau_a
\tau_b)=\delta_{ab} .
\]
This action is invariant under the usual Yang-Mills transformations
$\d_\om$ defined as
\eq\ba{l}
\d_\om A_\m = D_\m \om \es
\d_\om B_{\m\n\r} = [B_{\m\n\r},\om]
\ea\eqn{yangmills}
with $D_\m \cdots=\pa_\m\cdots + [A_\m, \cdots ]$. Furthermore, it
is also invariant under   the so-called reducible transformations
$\d_\p$ defined as
\eq\ba{l}
\d_\p A_\m = 0 \es
\d_\p B_{\m\n\r} = -(D_\m\p_{\n\r}+{\rm cyclic\ permutations}).
\ea\eqn{reductible}
where $\om$ and $\p$ are forms of degree 0 and 2 respectively.

As usual, one has to fix the gauge. The first point is to determine
the number of degrees of freedom of the field $B$.
Using \equ{reductible}, one sees that all the $\p$ which can be
written as\footnote{The action of the covariant derivative over a
form $\OM$ is given by $D\OM=dx^\m D_\m\OM $.}
$\p=D\p'$ where $\p'$ is a 1-form, are irrelevant on shell, i.e.
for fields solutions of their equation of motion. Indeed
the transformation \equ{reductible} then reads
$$
\d_\p B_{\m\n\r} = - ( F_{\m\n}\p'_\r +
                        {\rm cyclic\ permutations})
$$
and $F_{\m\n}=0$ is the field equation for $A_\m$.
By repeating the same argument for $\p'=D\p''$, where
$\p''$ is now a 0-form, one obtain that the number of degrees of
freedom of $B$ we have to fix is equal to $10-(5-1)=6$. On the
other hand, the field $A$ has one gauge degree of freedom to be
fixed, as usual.

The choice of the gauge fixing condition is naturally related to the
geometry of the problem under interest. The space-time  we will
consider here is $\real^5$, and
the boundary $\BB$ is the plane defined by $x^4=0$. Therefore, the
axial gauge
\eq\ba{l}
n^\r A_\r=0, \es
n^\r B_{\m\n\r}=0
\ea\eqn{gaugefix}
with $n^\m=(0,0,0,0,1)$ is the natural choice\footnote{The transverse
coordinates with respect to $\BB$ are denoted by $\xtr$ and are
labelled by $m,n,p,\cdots =0,1,2,3$.}.

The gauge fixed action then reads
\eq\ba{ll}
\Sigma =\tr \dint d^5x&\left\{ \tilde{B}^q\lp\pa_4 A_q-\pa_q A_4
+\lc A_4, A_q \rc \rp + \right. \es
 &\ \left. +\frac{1}{2}\e^{mnpq}B_{mn4}\lp\pa_p A_q + A_p A_q\rp +
\pi A_4+\frac{1}{2}\pi^{mn}B_{mn4} \Bigr\} \right.
\ea\eqn{act-tot}
where $\tilde{B}^q$ is the
four-dimensional dual of $B_{mnp}$ defined as
$$
\tilde{B}^q=\frac{1}{3!}\e^{qmnp}B_{mnp}
$$
together with $\e^{qmnp4}=\e^{qmnp}$.
We have not introduced the Faddeev-Popov
ghosts since, in the axial gauge, they are decoupled and thus they are
not needed.

It is easy to check that the gauge fixing given above is
sufficient, the Lagrange multiplier fields $\pi$, $\pi^{mn}$ fix the
seven gauge degrees of freedom we have counted above.

The full gauge fixed
action \equ{act-tot} still possesses invariances.
The gauge fixing terms
are responsible for the breaking of \equ{yangmills}, \equ{reductible}
as well as for the breaking of the
five-dimensional Poincar\'e invariance.
Nevertheless, we stay with
four-dimensional Poincar\'e invariance in the
coordinates transverse to the boundary,
together with two residual gauge invariances whose
transformation laws have the same form
as \equ{yangmills} and
 \equ{reductible} but where the gauge parameters do
not depend of $x^4$:
\eqa
\om&=&\om(x^0,x^1,x^2,x^3), \label{residual1} \es
\p_{mn}&=&\p_{mn}(x^0,x^1,x^2,x^3). \label{residual2}
\eea

Let us now introduce the boundary.
Following the procedure given in~\cite{emery-piguet}, the effects of
the boundary $\BB$ are specified by two conditions. The first one is
a decoupling condition which states that $\BB$
separates our space $\real^5$ in two half-spaces
labeled by $+$ and   $-$
corresponding to the sign of the $x^4$ component.
We also impose a locality condition which states that the behavior
away from the boundary is the same as the one of the theory
without boundary.

The simplest way to fill these conditions, \ie to describe the
effect of the boundary, is by working directly at the level of the
generating functional of the connected Green functions
$\zc[J^q,\tilde{J}_q,J_4,J^{mn4},J_\pi,J_{\pi^{mn}}]$, whose
arguments are the sources of the fields
$A_q$, $\tilde{B}^q$, $A_4$, $B_{mn4}$, $\pi$ and
$\pi^{mn}$ respectively.
Indeed, the decoupling condition corresponds to the decomposition
of $\zc$ into two parts
\eq
\zc(J_\vf)={\zc}_+(J_\vf)+{\zc}_-(J_\vf) \ ,
\eqn{decomp}
which implies that
an $n$-point Green function will be written as
\eq
\vev{\vf_{i_1}(x_1)\cdots\vf_{i_n}(x_n)} =
\th_+\vev{\vf_{i_1}(x_1)\cdots\vf_{i_n}(x_n)}_+ +
\th_-\vev{\vf_{i_1}(x_1)\cdots\vf_{i_n}(x_n)}_-
\eqn{$n$-point-g-fonction}
with
$$
\th_\pm=\th (\pm x_1^4)\th (\pm x_2^4)\cdots \th (\pm x_n^4)
$$
In particular the
propagators take the form
\eq
\Delta^{\vf \vf'}(x,x') = \th_+\D_+^{\vf \vf'}(x,x')
   +  \th_-\D_-^{\vf \vf'}(x,x')
\eqn{propdec}
where   $\theta_\pm=\theta(\pm x^4)\theta(\pm x'^4)$.

Then, the locality condition allows us to calculate the Green functions
$\vev{\cdots}_+$ and $\vev{\cdots}_-$  as the solutions
of the equations of motion of the theory without boundary.

The locality condition implying that
the effects of the boundary are local, the latter will be
described by terms in $\d(x^4)$. These terms will be
constrained by dimensional --
we want to preserve scale invariance --
and other symmetry arguments.
Therefore, the equations of motion of the theory
with boundary will be of the form
\eq\ba{ll}
-\pa_4\tilde{B}^q +\lc J_\pi,\tilde{B}^q\rc
 +\e^{mnpq}\pa_p J_{\pi^{mn}}
+\frac{1}{2}\e^{mnpq}\lc A_p,J_{\pi^{mn}}\rc +\!J^q \!\!&\!\!\!=
         \delta (\pm x^4)\lambda_\pm\tilde{B}^q_\pm  \\[2mm]
\pa_4 A_q-\lc J_\pi ,A_q\rc+\pa_q J_\pi+\tilde{J}_q \!\!&\!\!\! =
                \delta (\pm x^4)\lambda_\pm A_{q_\pm} \\[2mm]
\pa_q \tilde{B}^q+\lc A_q,\tilde{B}^q\rc+\pi+J^4
                      \!\!&\!\!\! = 0\\[2mm]
\e^{mnpq}\lp \pa_p A_q + A_p A_q\rp +\pi^{mn} + J^{mn4}
                       \!\!&\!\!\!  =0 \\[2mm]
A_4 + J_\pi    \!\!&\!\!\!= 0\\[2mm]
B_{mn4}+J_{\pi^{mn}}  \!\!&\!\!\!= 0
\ea\eqn{motion}
These equations have been written in a functional way,
with the notation
\eq
\vf(x) = \fud{\zc}{J_\vf(x)}\ ,
           \qquad \vf = A_m,\tilde{B}^q,A_4,B_{mn4},\pi,\pi^{mn}
\eqn{chpclass}
and where
$\vf_\pm(z)$ means the insertion of the field $\vf(x)$ on  the
right, respectively on the left of the boundary $\BB$:
\eq
\vf_\pm(z) = \lim_{x^4\to\pm 0} \fud{\zc}{J_\vf(x)}\ .
\eqn{fipm}

The last two equations of \equ{motion} are
the gauge conditions \equ{gaugefix} written in term of $\zc$.

Let us remark that the parameter $\la_\pm$ of the boundary terms
in the right-hand
sides of the first two field equations has been set equal for
both in order to assure their mutual consistency.
A motivation for the form of these boundary terms
may be found in the remark
that they could formally
be inferred from a surface term in the action of the
form
\eq
\tr\int_\BB d^4\xtr\, \tilde{B}^q A_q  =
 \tr\int d^5x\, \d(x^4)\tilde{B}^q A_q
\eqn{surfterm}
The arbitrariness of the coefficients $\la_\pm$ would then follow from
ambiguities caused by the multiplication of distributions
at the same point.

The gauge invariances \equ{yangmills}, \equ{reductible} of
the theory with boundary
also lead to functional identities. These identities,
due to the decoupling of the
Faddeev-Popov ghosts, take the form of two local Ward
identities:
\eqa
&&\displaystyle{\sum_{\varphi }}[J_\varphi ,  \varphi ]
-\pa_q J^q -\pa_4J^4-\pa_4\pi
   =-\d (\pm x^4)\la_\pm \pa_q\tilde{B}^q_\pm \label{wardloc1}\es
&&\ba{c} [J^{mn4} ,A_4 ]-\e^{mnpq}[\tilde{J}_q ,A_p ]
           + [J^\pi ,\pi^{mn} ]
+\e^{mnpq}\pa_p\tilde{J}_q -\pa_4\pi^{mn}- \es
- \pa_4 J^{mn4}
= -\d (\pm x^4)\la_\pm \e^{mnpq}\lp \pa_p A_{q_\pm}
+\frac{1}{2}\lc A_{p_\pm} ,A_{q_\pm}\rc\rp \ea  \label{wardloc2}
\eea


\section{The Free Propagators}  \label{prop.-libres}

We apply now the procedure described above to the computation
of the propagators. The locality condition implies the results shown
in Table \equ{prop1}.
\begin{table}
$$\lp\ba{cccc}
\ba{l} \d_{pq}\D^1_1\\
       + \pa_p\pa_q \D^1_2 \ea
& \ba{l} T_\x(x',x)\d_q^p \\
  +\pa_q\pa^p\D^2_1 \ea
& \ba{l} \pa_qT_\x(x',x) \\
  +\pa_q\pa^2\D^2_1 \ea
& \ba{l}\e^{rstp}\pa_t \D^1_1\d_{pq}\\$\ $\ea  \\
&&&\\
\ba{l} T_\x(x,x')\d^q_p \\
  +\pa_p\pa^q\D^2_1 \ea
&\ba{l}  \d^{pq}\D^3_1\\
       + \pa^p\pa^q \D^3_2 \ea
& \ba{l}\pa^q \D^3_1 \\
  +\pa^q \pa^2\D^3_2 \ea
& \ba{l}\e^{rstq}\pa_q\lac T_\x (x,x') \d_t^p\right.\\
  +\left. \pa_t\pa^p\D^2_1 \rac \ea  \es
&&&\\
\ba{l} -\pa_p T_\x(x,x') \\
  -\pa_p\pa^2\D^2_1 \ea
&  \ba{l}-\pa^p \D^3_1 \\
  -\pa^p \pa^2\D^3_2 \ea
&  \ba{l}-\pa^2 \D^3_1 \\
  -\pa^2 \pa^2\D^3_2 \ea  &\ba{l}0\\$\ $\ea  \es
&&&\\
\ba{l}\e^{mnrq}\pa_q \D^1_1\d_{rp}\\$\ $\ea
& \ba{l}\e^{mnqr}\pa_q\lac T_\x (x',x) \d_r^p\right. \\
\left. +\pa_r\pa^p \D^2_1  \rac \ea &\ba{l}0\\$\ $\ea
    & \ba{l}\e^{mnpq}\e^{rstl}\pa_p\pa_l \D^1_1 \d_{tq}\\$\ $\ea
\ea\rp
$$
\caption{\it The propagators   $\D_\pm(x,x')$. The table is
ordered according to the sequence
$A_{q(p)},\tilde{B}^{q(p)},\pi,\pi^{mn(rs)}$ for the columns
(resp.lines). The gauge indices have been dropped out since the
propagators are diagonal in the group space.}
\label{prop1}
\end{table}
We have set
$$
T_\x(x,x')=\lc \th (x^4-{x'}^4)+\x\rc\d^{(4)}(\xtr - x'{}^{{\rm tr}})
$$
and $(\x, \D^1_1, \D^1_2, \D^2_1, \D^3_1, \D^3_2 )$ are just
integration ``constants''. Note that, whereas $\x$ is just a number,
the $\D^i_j$ are arbitrary functions of the transverse coordinates,
more precisely of $(\xtr -x'{}^{{\rm tr}})^2$  due to the
four-dimensional
Poincar\'e invariance.

However, further restrictions on the propagators will
follow from  the decoupling condition.
Indeed, the form \equ{propdec}, due to the presence of
the $\th$-functions, will generate a boundary term when
substituted into \equ{motion}. Therefore, at the limit
$x^4\rightarrow \pm 0$, one gets the following constraints
between the parameters $\la_\pm $ and the integration constants:
\eq
\ba{ll}
\lp 1 -\la_+ \rp \D^1_1 =0          & \lp 1 +\la_- \rp \D^1_1 =0 \\
\lp 1 -\la_+ \rp \D^1_2 =0          & \lp 1 +\la_- \rp \D^1_2 =0 \\
\lp 1 -\la_+ \rp \D^2_1 =0          & \lp 1 +\la_- \rp \D^2_1 =0 \\
\lp 1 -\la_+ \rp \lp 1+\x\rp =0 \qquad& \lp 1 +\la_- \rp \x =0\\
\lp 1 +\la_+ \rp \D^3_1 =0         & \lp 1 -\la_- \rp \D^3_1 =0 \\
\lp 1 +\la_+ \rp \D^3_2 =0          & \lp 1 -\la_- \rp \D^3_2 =0 \\
\lp 1 +\la_+ \rp \D^2_1 =0          & \lp 1 -\la_- \rp \D^2_1 =0 \\
\lp 1 +\la_+ \rp \x =0        & \lp 1 -\la_- \rp \lp 1+\x\rp =0 \ .
\ea\eqn{constr}
The system \equ{constr} admit two solutions for each side of $\BB$:
\eq\ba{rllll}
{\rm I_+:} \ & \la_+=-1 \ & \x=-1 \ & \D^1_1=\D^1_2=\D^2_1=0
                \ & \D^3_1,\ \D^3_2\ {\rm arb.}\\
{\rm I_-:} \ & \la_-=+1 \ & \x=0 \ & \D^1_1=\D^1_2=\D^2_1=0
                \ & \D^3_1,\ \D^3_2\ {\rm arb.}\\
{\rm II_+:} \ & \la_+=+1 \ & \x=0 \ & \D^3_1=\D^3_2=\D^2_1=0
                \ & \D^1_1,\ \D^1_2\ {\rm arb.}\\
{\rm II_-:} \ & \la_-=-1 \ & \x=-1 \ & \D^3_1=\D^3_2=\D^2_1=0
                \ & \D^1_1,\ \D^1_2\ {\rm arb.}
\ea\eqn{sol}
These solutions are labelled according to the boundary
conditions implied by \equ{constr} on the fields at $\BB$:
${\rm I_\pm}$ corresponds to Dirichlet conditions (i.e.
vanishing of the fields on the corresponding side of the boundary)
for $A_q$, $\pi^{mn}$ whereas  ${\rm II_\pm}$ corresponds to
Dirichlet conditions for $\tilde{B}^q$, $\pi$.

Finally, it follows directly from the gauge
conditions, i.e. from the last two equations
of \equ{motion}, that the Green functions containing
$A_4$ and/or $B_{mn4}$ are all zero, except
\eq\ba{l}
\D_{\pi A_4}(x,x')=-\th_\pm\d^{(5)}(x-x') \\
\D_{\pi^{mn} B_{mn4}}(x,x')=-\th_\pm\d^{(5)}(x-x') \ .
\ea\eqn{propdelt}

  Let us display, for further use, the propagators corresponding to
the solution ${\rm I}_+$:
\eq\lp\ba{cccc}
0  & -T(x,x')\d^p_q &  -\pa_q T(x,x') & 0 \es
&&&\\
-T(x',x)\d^q_p & \d^{pq}\D^3_1+\pa^p\pa^q\D^3_2
 & \pa^q \D^3_1  +\pa^q \pa^2\D^3_2
 & \e^{rspq}\pa_q T(x',x)\es
&&&\\
\pa_p T(x',x)  &  -\pa^p \D^3_1  -\pa^p \pa^2\D^3_2
               &  -\pa^2 \D^3_1  -\pa^2 \pa^2\D^3_2  & 0 \es
&&&\\
0 & \e^{mnqp}\pa_q T(x,x') & 0 & 0
\ea\rp\eqn{prop+1}
where $T(x,x')=\th(x^4-x'^4)\d^{(4)}(\xtr-\xtr{}')$ and the
conventions are the same as in Table \equ{prop1}.
The integration ``constants'' $\Delta^3_1$, $\Delta^3_2$ are arbitrary
function of the transverse coordinates $(\xtr-\xtr{}')^2$.


\section{General Solution}\label{sol.-generale}

We want now to discuss the general solution of the field
  equations~\equ{motion}
for the $n$-point Green functions of the theory.
We shall restrict our analysis starting from the solution $I$
for the free propagators, an analogous discussion been possible
starting from solution II. We choose the
solution ${\rm I}_+$, the physical implications of ${\rm I}_-$ being
identical. In this context, the equations of motion
are given by \equ{motion}, with $\lambda_+=-1$.

\begin{figure}
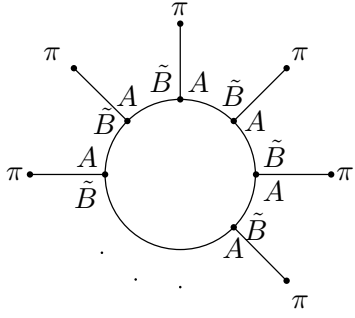

\begin{center}

\begin{texdraw}
\drawdim cm
\linewd 0.02
\textref h:C v:C
\move(4 3) \lcir r:1
\move(3 3) \fcir f:0 r:0.04 \lvec(2 3) \fcir f:0 r:0.04
            \htext(1.8 3){\small $\pi$}
\move(5 3) \fcir f:0 r:0.04 \lvec(6 3) \fcir f:0 r:0.04
           \htext(6.2 3){\small $\pi$}
\move(4 4) \fcir f:0 r:0.04 \lvec(4 5) \fcir f:0 r:0.04
           \htext(4 5.2){\small $\pi$}
\move(4.707 3.707) \fcir f:0 r:0.04 \lvec(5.414 4.414) \fcir f:0 r:0.04
           \htext(5.6 4.6){\small $\pi$}
\move(3.293 3.707) \fcir f:0 r:0.04 \lvec(2.586 4.414) \fcir f:0 r:0.04
           \htext(2.3 4.6){\small $\pi$}
\move(4.707 2.293) \fcir f:0 r:0.04 \lvec(5.414 1.586) \fcir f:0 r:0.04
           \htext(5.6 1.3){\small $\pi$}
\move(4 1.5) \fcir f:0 r:0.02
\move(2.9535 1.9535) \fcir f:0 r:0.02
\move(3.4 1.6) \fcir f:0 r:0.02
\textref h:L v:B \htext(4.1 4.1){\small $A$}
\textref h:R v:B \htext(3.9 4.1){\small $\tilde{B}$}
\textref h:C v:B \htext(3.293 3.9){\small $A$}
\textref h:R v:C \htext(3.15 3.707){\small $\tilde{B}$}
\textref h:R v:B \htext(2.9 3.1){\small $A$}
\textref h:R v:T \htext(2.9 2.9){\small $\tilde{B}$}
\textref h:C v:B \htext(4.707 3.9){\small $\tilde{B}$}
\textref h:L v:C \htext(4.85 3.707){\small $A$}
\textref h:L v:B \htext(5.1 3.1){\small $\tilde{B}$}
\textref h:L v:T \htext(5.1 2.9){\small $A$}
\textref h:L v:C \htext(4.85 2.3){\small $\tilde{B}$}
\textref h:C v:T \htext(4.707 2.15){\small $A$}
\end{texdraw}

\end{center}
\caption{\it Loop contributions to the Green functions of the field $\pi$.}
\label{figloop}
\end{figure}

As in~\cite{emery-piguet}, the third and fourth equations
  in~\equ{motion} could suffer
from ill-defined products of fields. More precisely, they
may generate divergent loops through the nonlinear terms
$[A_q,{\tilde B}^q]$ and $\e^{mnpq}A_pA_q$.
These diagrams would contribute to the Green functions of
the Lagrange multiplier field\footnote{The vanishing of
the propagator $\vev{AA}$ for this solution forbids
the existence of loops for $\pi^{mn}$.} $\pi$.
Nevertheless, these {\it a priori} UV divergent contributions shown
in Fig.~\ref{figloop} factorize into a divergent
$x^4-$independent part
which can be regularized~\cite{emery-piguet},
and an $x^4-$dependent part of the form
$\th(x_1^4 - x_2^4)\th(x_2^4 - x_3^4)\cdots \th(x_n^4 - x_1^4)$
which is zero by itself. This   shows the absence of any radiative
corrections and thus allows one to neglect the nonlinear terms in the
third and fourth equations \equ{motion}.

\begin{figure}
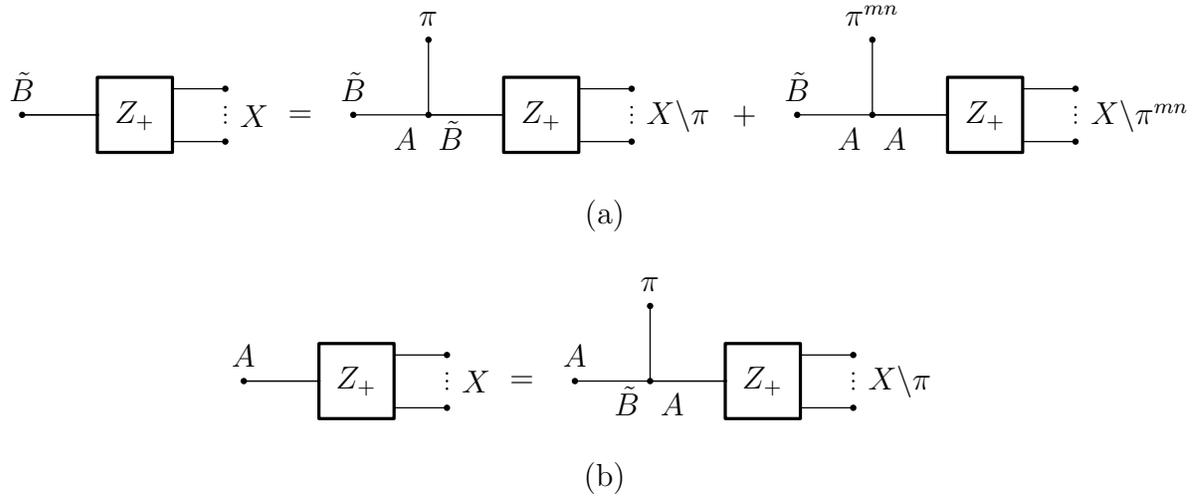

\begin{center}
\begin{texdraw}
\drawdim cm
\linewd 0.02

\move(1 2)   \mayo \rmove(0.2 -0.1)\textref h:L v:C \htext{$X$}

\textref h:C v:C \htext(4.7 2){$=$}

\move(5.4 2) \fcir f:0 r:0.04 \rlvec(1 0)
              \rlvec(0 1) \fcir f:0 r:0.04
              \rmove(0 -1) \mayo
              \rmove(0.2 -0.1)\textref h:L v:C \htext{$ X \backslash \pi$}

\textref h:C v:C \htext(10.6 2){$+$}

\move(11.3 2) \fcir f:0 r:0.04 \rlvec(1 0)
              \rlvec(0 1) \fcir f:0 r:0.04
              \rmove(0 -1) \mayo
              \rmove(0.2 -0.1)\textref h:L v:C \htext{$X\backslash \pi^{mn}$}

\textref h:C v:B \htext(1 2.2){$\tilde{B}$}
                 \htext(5.4 2.2){$\tilde{B}$}
                 \htext(6.1 1.55){$A$}
                 \htext(6.7 1.55){$\tilde{B}$}
                 \htext(6.4 3.2){$\pi$}
                 \htext(11.3 2.2){$\tilde{B}$}
                 \htext(12  1.55){$A$}
                 \htext(12.6 1.55){$A$}
                 \htext(12.3 3.2){$\pi^{mn}$}
\vspace*{1cm}
\end{texdraw}

\begin{texdraw}
\drawdim cm
\linewd 0.02
\move(1 2)   \mayo \rmove(0.2 -0.1)\textref h:L v:C \htext{$X$}

\textref h:C v:C \htext(4.7 2){$=$}

\move(5.4 2) \fcir f:0 r:0.04 \rlvec(1 0)
              \rlvec(0 1) \fcir f:0 r:0.04
              \rmove(0 -1) \mayo
              \rmove(0.2 -0.1)\textref h:L v:C \htext{$X \backslash \pi$}
\textref h:C v:B \htext(1 2.2){$A$}
                 \htext(5.4 2.2){$A$}
                 \htext(6.1 1.55){$\tilde{B}$}
                 \htext(6.7 1.55){$A$}
                 \htext(6.4 3.2){$\pi$}
                 \htext(5.8 4){(a)}
                 \htext(5.8 0.5){(b)}
\end{texdraw}

\end{center}
\caption{\it Diagrammatic representation of $(4.8)$ {\rm (a)}
                 and $(4.9)$ {\rm (b)}.}
\label{figdiag}
\end{figure}

The first two equations   in~\equ{motion} give rise to recursion
relations for the Green functions. Such relations are generated by
differentiation with respect to the sources, \ie by means of a
general functional operator
\eq
\X= \frac{\d^{L+M+N+P}}{\lp\d J^A\rp^L\lp\d J^{\tilde{B}}\rp^M
           \lp\d J^{\pi} \rp^N\lp\d J^{\pi^{mn}} \rp^P}
\eqn{gen-op}
We begin the analysis outside from the boundary, i.e. we consider first
the equations for the component $\vev{\cdots}_+$ of the
decomposition \equ{$n$-point-g-fonction}. This gives
\eqa
\pa_4\vev{{\tilde B}^{q\,a}(x) X}_+\=
 \sum_{i=1}^N f^{a a_i e}\d^{(5)}(x-x_i)
 \vev{{\tilde B}^{q\,e}(x_i) X \backslash \pi^{a_i}(x_i)}_+
                                                 \label{recb}\es
 \+ \sum_{j=1}^P f^{a b_j e}\e^{m_j n_j pq}\d^{(5)}(x-y_j)
 \vev{A_p^e(y_j) X \backslash \pi^{m_j n_j\,b_j}(y_j)}_+
                                                            \nes
\+ \d_{X, A^a_q(y)}\d^{(5)}(x-y)
           +\d_{X,\pi^{mn\,a}(z)}\d^{(5)}(x-z) \nes
\pa_4\vev{A_q^{a}(x) X}_+\=
      \sum_{i=1}^N f^{a a_i e}\d^{(5)}(x-x_i)
      \vev{A_q^e(x_i) X \backslash  \pi^{a_i}(x_i)}_+
                                                      \label{reca}\nes
           \+ \d_{X, {\tilde B}^{q\,a}(y)}\d^{(5)}(x-y)
              + \d_{X,\pi^a(z)}\d^{(5)}(x-z)
\eea
  where
\eq
X = \lp\dprod{1}{L} A^{a_i}_{m_i}(w_i)\rp
    \lp\dprod{1}{M} {\tilde B}^{m_i\,b_i}(z_i)\rp
    \lp\dprod{1}{N} \pi^{c_i}(x_i)\rp
    \lp\dprod{1}{P} \pi^{m_in_i\,d_i}(y_i)\rp
\eqn{def-x}
and where $X\backslash \vf$ means
the omission of the field $\vf$   from
the string $X$. The solution for $L+M+N+P\geq 2$ has the form
of the following recursion relations:
\eqa
\lefteqn{\vev{\tilde{B}^{q\,a}(x) X}_+=-\sum_{i=1}^N f^{a c_i e}
        \th(x_i^4-x^4)\d^{(4)}(\xtr-\xtr_i)
 \vev{\tilde{B}^{q\,e}(x_i) X \backslash \pi^{c_i}(x_i)}_+}\nes
&&
\!\!-\!\!\sum_{j=1}^P f^{a d_j e}\e^{m_j n_j pq}
  \,\th(y_j^4-x^4) \d^{(4)}(\xtr-\ytr_j)
  \vev{A_p^e(y_j) X \backslash \pi^{m_j n_j\,d_j}(y_j)}_+ \nes
 && + \vev{\tilde{B}^{q\,a}(\xtr) X}_+  \label{rec-rel-1} \\[4mm]
\lefteqn{\vev{A_q^{a}(x) X}_+ =
   \sum_{i=1}^N f^{a c_i e}\,
   \th(x^4-x_i^4)\d^{(4)}(\xtr-\xtr_i)
      \vev{A_q^e(x_i) X \backslash \pi^{c_i}(x_i)}_+}  \nes
&&+ \vev{A_q^{a}(\xtr) X}_+ \nonumber
\eea
Bose statistics and the consistency of
the procedure -- a same Green
function can be determined in various ways by the recursion, and the
2-point function $\vev{A(x)\tilde B(x')}_+$ {\it does} depend on
$x^4$ and ${x'}^4$ --
fix the ``integration constants'' $\vev{\tilde{B}^{q\,a}(\xtr) X}_+$
and $\vev{A_q^{a}(\xtr) X}_+$
to zero, except for the case of the Green functions of the fields $A$
and $\tilde B$ alone (corresponding to $N=P=0$ in \equ{def-x}).
The only information that we have  for the latter up to this point is
that they depend only on the transverse coordinates. As we did for the
2-point functions, we have to take into account the effect of the
boundary. The complete  equations  for these Green
functions, following from the second of the field equations~\equ{motion},
read (with $\lambda_+=-1$)
\eq
\pa_{x^4}\vev{A(x) \dprod{1}{L}A(w^{\rm tr}_i)
  \dprod{1}{M}\tilde B(z^{\rm tr}_i)} =
  -  \d(x^4) \vev{A(x) \dprod{1}{L}A(w^{\rm tr}_i)
  \dprod{1}{M}\tilde B(z^{\rm tr}_i)}
\eqn{green-A-B}
whereas the decoupling condition expressed by \equ{$n$-point-g-fonction}
yields the same equation but with the opposite sign for the right-hand
side. On the other hand, use of the first of the field equations
\equ{motion} does not lead to a contradictory sign.
Hence the result
\eq\ba{l}
\vev{\dprod{1}{L}A(w^{\rm tr}_i)
  \dprod{1}{M}\tilde B(z^{\rm tr}_i)} = 0\quad\mbox{for}
    \quad L\not=0 \\[4mm]
\vev{\dprod{1}{M}\tilde B(z^{\rm tr}_i)} \quad\mbox{arbitrary}
\ea\eqn{ABandB}
Let us come back to the Green functions involving the Lagrange fields.
Taking into account the
expression \equ{prop+1} for the propagators of ${\rm I}_+$, we can write
the recursion relations for these Green functions in the following form:
\eqa
\vev{\tilde{B}^{q\,a}(x) X}_+\=
\sum_{i=1}^N f^{e_i c_i e} \vev{\tilde{B}^{q\,a}(x)A_{q_i}^{e_i} (x_i)}_+
\vev{\tilde{B}^{q_i\,e}(x_i) X \backslash \pi^{c_i}(x_i)}_+
                                      \label{dia1}\es
   \+ \!\!\sum_{j=1}^P \e^{m_j n_j pq}f^{e_j d_j e}
   \vev{\tilde{B}^{q\,a}(x)A_{q_j}^{e_j} (y_j)}_+
    \vev{A^e_{q_j} (y_j) X \backslash
                             \pi^{m_j n_j\,d_j}(y_j)}_+     \nes
\vev{A_q^a(x) X}_+\=
 -\sum_{i=1}^N f^{e_i c_i e}\vev{A_q^a (x) \tilde{B}^{q_i\,e_i}(x_i)}_+
   \vev{A^{e}_{q_i}(x_i) X \backslash \pi^{a_i}(x_i)}_+ \label{dia2}
\eea
and this allows for the diagrammatical representation
shown in Fig.~\ref{figdiag}.

The non-vanishing Green functions generated by this procedure
divide in two classes.

The first class is made of the Green functions of the type
\eq\ba{c}
\vev{(A)(\tilde B)(\pi)^N}\ ,\quad
\vev{(\tilde B)(\pi)^N(\pi^{mn})}\ ,
\quad
\vev{(\tilde B)^2(\pi)^N(\pi^{mn})}\ ,\es
\quad \vev{(A)(\pi)^N}\ ,
\quad \vev{(\tilde B)(\pi)^N}
 \qquad (N\ {\rm  arbitrary})
\ea\eqn{class-1}
One sees that they are completely determined by the recursion
relations and the two-point functions.
They correspond in fact to
the tree graphs generated by the Feynman rules defined
by the propagators~\equ{prop+1} and
the BF vertex read off from the action.
A typical example is shown in Fig.~\ref{figpiN}
\subsubsection*{Remark:}
\begin{figure}
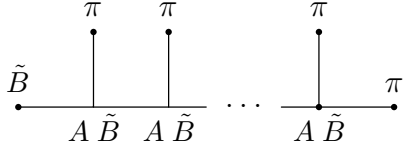

\begin{center}
\begin{texdraw}
\drawdim cm
\linewd 0.02
\move(1 2)
              \fcir f:0 r:0.04 \rlvec(1 0)
              \rlvec(0 1) \fcir f:0 r:0.04
              \rmove(0 -1) \rlvec(1 0)
              \rlvec(0 1) \fcir f:0 r:0.04
              \rmove(0 -1) \rlvec(0.5 0)
              \rmove(1 0) \rlvec(0.5 0)
              \fcir f:0 r:0.04 \rlvec(0 1) \fcir f:0 r:0.04
              \rmove(0 -1) \rlvec(1 0) \fcir f:0 r:0.04
\textref h:C v:B \htext(1 2.2){\small $\tilde{B}$}
                 \htext(1.8 1.55){\small $A$}
                 \htext(2 3.2){$\pi$}
                 \htext(2.2 1.55){\small $\tilde{B}$}
                 \htext(2.8 1.55){\small $A$}
                 \htext(3 3.2){$\pi$}
                 \htext(3.2 1.55){\small $\tilde{B}$}
                 \htext(4 2){$\ldots$}
                 \htext(4.8 1.55){\small $A$}
                 \htext(5 3.2){$\pi$}
                 \htext(5.2 1.55){\small $\tilde{B}$}
                 \htext(6 2.2){$\pi$}
\end{texdraw}

\end{center}
\caption{\it Diagrammatic representation
of  $\left\langle (\tilde B)(\pi)^N\right\rangle$.}
\label{figpiN}
\end{figure}

{\it As we will see in Section~\ref{alg.-courants},
the residual Ward identities will impose some constraints on these
  Green functions, more precisely a transversality condition for the
propagator $\vev{\tilde{B}\tilde{B}}$ (see \equ{transbb}).
One of the consequences
is that the propagator $\vev{\pi\tilde{B}}$ \equ{prop+1}
will vanish. Therefore,
all the Green functions of the type $\vev{(\tilde B)(\pi)^N}$
-- shown in Fig.~\ref{figpiN} --  will be zero, once the residual
gauge invariance is taken into account, since they involve this
propagator.}


The second class is made of the Green functions
\eq
\vev{(\tilde B)^M (\pi)^N}\ \qquad (M\ge2,\ N\ {\rm arbitrary})
\eqn{class-2}
For $N=0$, they are the arbitrary Green functions
of $\tilde{B}$ (see second line of \equ{ABandB}).
Those for $N\ge1$ are determined from the former through the
recursion relations \equ{dia1}. They correspond
to the tree graphs generated by the same set of Feynman rules,
but starting from a trunk
given by one of the Green functions $\vev{(B)^M}$.

A priori, we still have to examine the consequences
of the equations of motion
for the Lagrange   multiplier fields. Indeed,
the fourth of the field equations \equ{motion} allows to
compute directly $\vev{\pi^{mn}\vf\cdots\vf}$
in terms of Green functions with the field $\pi^{mn}$ replaced by a
derivative of the field $A$,  and the third one leads to a similar
dependance for $\vev{\pi\vf\cdots\vf}$. However, the latter Green
functions --   except the ones involving only the Lagrange
multiplier fields -- have already been generated
  by the recursion relations \equ{dia1} and \equ{dia2} following
from the first two equations \equ{motion}.
Therefore, we have to address the
problem of consistency between these two procedures. For this purpose,
let us rewrite \equ{wardres1} as a functional operator $W(x)$
\eq
W(x)Z_+\equiv\lac \pa_q\fud{}{\tilde{J}_q{}_+}
     -\int_0^{+\infty}\sum_{\vf}\lc J^\vf\fud{}{J^\vf}\rc\rac Z_+
\eqn{op-ward}
and, define the linearized operators $M^\pi$, $M^{\pi^{mn}}$
corresponding to the third and fourth equations \equ{motion} as
\eqa
M^\pi Z_+&\equiv&\lac \fud{}{J_\pi} + \pa_q \fud{}{\tilde{J}_q}
+\lc \fud{Z_+}{J^q},\fud{}{\tilde{J}_q} \rc +
\lc \fud{Z_+}{\tilde{J}_q} ,\fud{}{J^q} \rc  \rac Z_+ \nes
M^{\pi^{mn}} Z_+ &\equiv& \lac \fud{}{J_{\pi^{mn}}}
+ \e^{mnpq} \pa_p \fud{}{J^q}
+2\e^{mnpq} \lc \fud{Z_+}{J^p}, \fud{}{J^q}\rc\rac Z_+ \nonumber
\eea
Thus, it is easy to check that
\eq\ba{c}
\lc M^{\pi\,a}(x), W^b(y)\rc=0 \es
\lc M^{\pi^{mn\,a}}(x), W^b(y)\rc=0
\ea
\eqn{consistence}
which is nothing else than a consistency relation between the two
procedures. This concludes the analysis of the solution ${\rm I}_+$.


\section{Ward Identities and Current Algebra}\label{alg.-courants}
In order to get the Ward identities expressing the residual gauge
invariance of the theory,
i.e. the invariance under the gauge transformations \equ{yangmills} and
\equ{reductible},
one has to integrate the local Ward identities \equ{wardloc1},
\equ{wardloc2} over $x^4$.
This step suffers from a long distance problem inherent to the
choice of an axial gauge. Indeed, from a naive computation,
these residual Ward identities would have the form
\eqa
&\displaystyle{\int^{+\infty}_{-\infty}}dx^4 \lac\sum_{\varphi }
               [J_\varphi ,\varphi ]-\pa_q J^q \rac
 =-\la_\pm \pa_q\tilde{B}^q_\pm &
                                      \label{wardres1} \es
&\ba{r}\displaystyle{\int^{+\infty}_{-\infty}}dx^4\lac [J^{mn4} ,A_4 ]
-\e^{mnpq}[\tilde{J}_q ,A_p ]
+[J^\pi ,\pi^{mn} ] +\e^{mnpq}\pa_p\tilde{J}_q \rac = \es
  = - \la_\pm \e^{mnpq}\lp
     \pa_p A_{q_\pm} +
     \frac{1}{2}\lc A_{p_\pm} ,A_{q_\pm}\rc\rp \ea &
                                          \label{wardres2}
\eea
Comparing with the local Ward identities \equ{wardloc1}, \equ{wardloc2},
one sees that  the validity
of \equ{wardres1}, \equ{wardres2} is equivalent to the
conditions
\eqa
\displaystyle{\int^{+\infty}_{-\infty} dx^4\, \pa_4
              \vev{\pi\vf \cdots \vf }=0} \label{zeromode2}\es
\displaystyle{\int^{+\infty}_{-\infty} dx^4\, \pa_4
              \vev{\pi^{mn}\vf \cdots \vf}=0}    \label{zeromode3}
\eea
However these conditions
are not fulfilled for neither of the solutions I or II discussed
in Section~\ref{prop.-libres}. This is essentially due to the fact that,
for any of these solutions, at least some of the propagators
$\vev{\vf(x)\vf'(x')}$   do not vanish at
infinite $x^4$. For the solution ${\rm I}_+$, for instance (see
\equ{prop+1}), this is the case for $\vf(x)=A(x)$ and for
$\vf(x)=\pi^{mn}(x)$, and it is not difficult to see,
by examining examples involving low-point functions, e.g.
$\vev{\pi A\tilde B}$
or $\vev{\pi^{mn}\tilde B}$, that this leads to a
violation of both conditions \equ{zeromode2}, \equ{zeromode3}.

In order to cure this pathological feature of the axial gauge, we may
introduce an infrared regularization. It turns out that an appropriate
way is to replace the gauge terms in \equ{act-tot}
by
\eq
\S^\e_{\rm gf} = \tr\dint\ d^5x\, e^{\e (x^4)^2}\lac \pi A_4
   + \dfrac{1}{2}\pi^{mn}B_{mn4}\rac
\eqn{reg-gf}
where the IR cut-off $\e$ is a positive number.
The resulting modification of the field equations \equ{motion}
consists of the substitutions
\eq\ba{ll}
\pi\to e^{\e (x^4)^2} \pi\qquad&\pi^{mn}\to e^{\e (x^4)^2}\pi^{mn}\es
J_\pi\to e^{-\e (x^4)^2}J_\pi\qquad
     &J_{\pi^{mn}}\to e^{-\e (x^4)^2} J_{\pi^{mn}}
\ea\eqn{reg-fields-sources}
This leads to a damping factor $\e^{-N\e (x^4)^2}$
in front of each Green function, where $N$ is its number of fields
$\pi$ and $\pi^{mn}$. This is sufficient for guarantying the validity of
the conditions \equ{zeromode2}, \equ{zeromode3}, hence of the residual
gauge Ward identities \equ{wardres1} and \equ{wardres2}. We don't
need to write
the latter again since they don't depend explicitly on $\e$.
\subsubsection*{Remarks:}
\begin{itemize}
{\it \item[1.] The introduction of the infrared regularization
does not change
anything to the discussion and to the conclusions of
Sections~\ref{prop.-libres} and \ref{sol.-generale}.
\item[2.]
Since the infrared cut-off $\e$ appears
in the gauge fixing term, one expects the physical quantities not to
depend on it. This will actually be the case for the current Green
functions to be defined below.
}\end{itemize}
Let us now examine the consequences of the residual Ward identities, for
each of the solution I and II, keeping ourselves on the + side.
\subsection*{Solution ${\rm I}_+$}
The Ward identity \equ{wardres1}, which in terms of Green functions,
reads
\eq
\int_0^{+\infty} dx^4\lac \sum_{\vf}
    f^{abc}\d^{(5)}(x-y)\vev{\vf^c(y)X\backslash \vf^b(y)}^\e\rac
      =\left. \pa_q \vev{\tilde{B}^{q^a}(x) X}^\e\right\vert_{x^4=+0}
\eqn{recwar}
-- the upperscript $\e$ reminds one that the theory is
now regularized --
gives essentially restrictions  to the
Green functions of the field $\tilde{B}$.
For those involving only $\tilde{B}$, this gives the
equations
\eq\ba{l}
\pa_q\vev{\tilde{B}^{q\,a} (\xtr ) \tilde{B}^{q_1\,{a_1}} (\xtr_1) \cdots
\tilde{B}^{q_N\,{a_N}} (\xtr_N)}= \es +\dsum{i=1}{N}f^{a a_i b}
\d^{(4)}(\xtr-\xtr_i) \vev{\tilde{B}^{q_i\,b}
(\xtr)\tilde{B}^{q_1\,{a_1}} (\xtr_1) \ldots
   \widehat{\tilde{B}^{q_i\,{a_i}} (\xtr_i)}
   \ldots \tilde{B}^{q_N\,{a_N}} (\xtr_N)}
\ea\eqn{recward1}
  where the hat on an argument means the omission of the latter.
We have taken into account the fact that the Green functions of $\tilde{B}$
depend only on the transverse coordinates, and we recall that they do
not depend on the infrared cut-off.

For $N=1$, in particular, we get the transversality condition
\eq
\pa_q\vev{\tilde{B}^q\tilde{B}^p}=0
\eqn{trans-2-pt}
which implies (see \equ{prop+1})
\eq
\D^3_1=-\pa^2 \D^3_2
\eqn{transbb}
and, subsidiarily, the vanishing of the propagators $\vev{\pi\tilde{B}}$ and
$\vev{\pi\pi}$.

Now, defining the currents on the + side of the boundary $\BB$ as
\eq
V^{p\,a}(\xtr)=\lim_{x^4\rightarrow +0} \tilde{B}^{p\,a} (x),
\eqn{def-currents}
the Ward identities \equ{recward1} imply the conservation law
\eq
\pa_q V^{q\,a}(\xtr)=0
\eqn{curr-cons}
and the equal-time current algebra
\eq
\lc V^{0\,a} (\xtr),V^{p\,b} (x'{}^{\rm tr}) \rc_{x_0=y_0}
= f^{abc}\d^{(3)}({\bf x}-{\bf x}')V^{p\,c} (\xtr)
\eqn{algB}

Turning now towards the  Ward identity \equ{wardres2} we see that
its right-hand side vanishes since the solution considered
here corresponds to a Dirichlet condition for the field $A$. It
therefore does not gives new information
concerning the physics on $\BB$.
\subsection*{Solution ${\rm II}_+$}
This solution corresponding to a Dirichlet condition for the field
$\tilde{B}$, it is the Ward identity \equ{wardres1}
which now becomes empty on the boundary.

On the other hand, the hard breaking
of \equ{wardres2} caused by the presence of the boundary prevents us to
interpret it as a current algebra.
One sees actually that its interpretation,  on the
boundary, is simply the equation
\eq
F_{mn}= 0,
\eqn{Fmn=0}
i.e. the vanishing of the Yang-Mills strength.
This means that this solution gives a four-dimensional
theory of topological type.


\section{Conclusion}

We have thus shown that, of the two possible solutions
${\rm I_\pm}$ and ${\rm II_\pm}$ of the field
equations \equ{motion}, only one, namely ${\rm I_\pm}$,
generates a current algebra
on $\BB$ \equ{algB}. This current algebra follows from
the Ward identity describing the residual gauge invariance of
the Yang-Mills type. The other residual Ward identity
becomes empty, on $\BB$,
due to the Dirichlet boundary condition for the field $A$.

On the other hand, the solution ${\rm II_\pm}$ generates
from the second residual Ward identity an identity
which only reproduces on the boundary
the field equation $F_{mn}=0$, which was already
known to hold outside from the boundary.
It is the Yang-Mills residual Ward identity
which is empty, in this case.

The infrared cut-off $\e$ needed for the validity of the residual gauge
invariance Ward identities does not affect the theory on the boundary
which constitutes the physical output of the present
considerations.

Finally, the presence of the boundary also
leads to a theory which is free of
radiative corrections. Indeed, it fixes the value of the
integration constants in such a way that the only possible
loops, which appear in the Green functions of the Lagrange multiplier
field $\pi$, are products of $\th$-functions which close on themselves.
Thus they are zero.

\subsection*{Acknowledgement}

One of the us (S.E.) would like to thank Prof. M. Schweda and all the
members of the Institut f\"ur Theoretische Physik of the Technische
Universit\"at Wien for their warm hospitality.

\end{document}